%% file: main.tex
\documentclass[%
preprint,
 amsmath,amssymb,
 aps,
]{revtex4-2}

\usepackage{graphicx}
\usepackage{dcolumn}
\usepackage{bm}

\usepackage{verbatim}

\usepackage{hyperref} 

\usepackage[normalem]{ulem}
\usepackage{color}

\newcommand{\AL}[1]{\textcolor{black}{#1}}

\newcommand{\CRCA}[1]{\textcolor{black}{#1}}

\newcommand{\CRCAC}[1]{\textcolor{black}{#1}}

\input{MDGcommands}

\begin{document}

\title{Data-driven prediction of large-scale spatiotemporal chaos with distributed low-dimensional models}
\author{C. Ricardo Constante-Amores$^{1,2}$}
 \email{crconsta@illinois.edu}
\author{Alec J. Linot$^{3}$}
\author{Michael D. Graham$^{1}$}%
\affiliation{$^1$Department of Chemical and Biological Engineering,
University of Wisconsin-Madison,
Madison, WI 53706, USA \\
$^2$Department of Mechanical Science and Engineering, University of Illinois, Urbana Champaign, IL 61801, USA\\
$^3$Department of Mechanical and Aerospace Engineering, University of California, Los Angeles, CA 90095, USA
}

\begin{abstract}

Complex chaotic dynamics, seen in natural and industrial systems like turbulent flows and weather patterns, often span vast spatial domains with interactions across scales. Accurately capturing these features requires  a high-dimensional state space
to resolve all the time and spatial scales. For dissipative systems  the dynamics  lie on a finite-dimensional manifold with fewer degrees of freedom. 
Thus, by building reduced-order data-driven models in manifold coordinates, we can capture the essential behavior of \CRCAC{chaotic} systems.
Unfortunately, these tend to be formulated globally rendering them less effective for large spatial systems. In this context, we present a data-driven low-dimensional modeling approach to tackle the complexities of chaotic motion, Markovian dynamics, multi-scale behavior, and high numbers of degrees of freedom within large spatial domains.  Our methodology involves a parallel scheme of decomposing a spatially extended system into a sequence of local ``patches'', and constructing a set of coupled, local low-dimensional dynamical models for each patch.
Here, we choose to construct the set of local models using autoencoders (for constructing the low-dimensional representation) and neural ordinary differential equations, NODE, for learning the evolution equation. 
\CRCAC{Each patch, or local model, shares the same underlying functions (e.g., autoencoders and NODEs)} \AL{due to the spatial homogeneity of the underlying systems we consider.}
\AL{We apply this method to the Kuramoto-Sivashinsky equation and 2D Kolmogorov flow, and reduce state dimension by up to two orders of magnitude while accurately capturing both short-term dynamics and long-term statistics.}

\end{abstract}

\maketitle

\section{Introduction}

Many important processes in nature are characterized by chaotic motion and exhibit structures across various spatial and temporal scales. Examples of this include weather, climate, turbulent flows, ocean dynamics \CRCA{and chemical reactions} \cite{lau2011intraseasonal,pope2001turbulent,Rinzel}.
Turbulent flows, in particular, serve as a prime illustration of how local interactions within a vast spatial domain intricately shape global behavior across multiple scales \cite{sagaut2008homogeneous}. While our main focus in this work centers on \CRCA{ the 1D  Kuramoto-Sivashinsky Equation (KSE) and 2D turbulent Navier-Stokes Equations (NSE),} 
the ensuing framework applies to any large spatial domain systems whose dynamics are governed by dissipative PDEs, where the long-time dynamics lie on a finite-dimensional manifold, and can be represented by a network of local interactions.

\CRCAC{The study of turbulence, particularly in wall-bounded flows, is essential for advancing our understanding of fluid dynamics, given its significant implications for various engineering applications \cite{pope2001turbulent,avila}. Approximately $25\%$ of the energy consumed by industry is due to transporting fluids through pipes and channels, with about one-quarter of this energy lost because of turbulence near walls \cite{jimenez,Waleffe,Jimenez_2013}. Due to the critical role of turbulence in energy dissipation, developing low-dimensional models for turbulent flows remains a formidable challenge. 
}
The advent of high-performance computing has resulted in the availability of large datasets, propelling a myriad of research efforts focused on building data-driven Reduced-Order Models (ROMs)  \cite{Taira,Lee,Eivazi,Schmid}. \AL{These ROMs facilitate fast forecasting, control, and interpretability of turbulent systems \cite{rowley_2009,Page_2021,alec_coutte,fukami2023grasping,pipe_jfm}.}

\CRCAC{Solutions of the NSE are formally infinite-dimensional. When simulating we approximate them on a high-resolution grid to capture all the relevant scales. However, due to viscosity the solutions actually lie on an (invariant) manifold of much lower dimension than the high-resolution simulation \cite{Titi,Foias,Hopf}. Thus, we should be able to reduce the dimension much more dramatically.}
The pursuit of \AL{low-dimensional representations} of these invariant manifolds for turbulent problems has led to a variety of data-driven frameworks, as shown by \cite{carlos,alec_coutte,Page_2021,Kaszas}.

\AL{One of the most popular methods of}
dimension reduction \CRCA{and manifold dimension estimation} from data is principal components analysis  (PCA), a linear technique that projects system states onto an orthogonal basis, organized by their contribution to the variance \cite{Jolliffe1986,abdi2010principal}.
\CRCAC{PCA, being a linear reduction method, projects data onto a flat manifold, which is not generally appropriate for} \AL{a minimal representation of }\CRCA{complex nonlinear problems} \cite{Floryan}.
\CRCAC{ Thus, nonlinear methods can be advantageous. Autoencoders are one of the most prevalent methods for nonlinear dimension reduction.}
They are a pair of neural networks in which one network maps a high-dimensional space to a low-dimensional space, and the other maps back \citep{Kramer,Hinton,Milano}.
\CRCA{A recently developed IRMAE-WD (implicit rank-minimizing autoencoder-weight decay) autoencoder architecture is capable of identifying \AL{a coordinate system for the manifold with the correct manifold dimension}
from data, for complex nonlinear dynamical systems \cite{jing2020implicit, irmae,koopman_pipe}.}

\AL{With a low-dimensional coordinate representation of the manifold, we can next discover an evolution equation in this coordinate system.}
Consider first the full-state data $u$ that live in ambient space $u\in \mathbb{R}^d$, where $du/dt=f(u)$ governs the time-evolution of this state. 
When the full space is mapped into the coordinates of a low-dimensional space, where $h\in \mathbb{R}^{d_h}$, an equivalent evolution equation 
in \CRCA{manifold} coordinates can be expressed as $dh/dt=g(h)$. 
The aim of the data-driven modeling is to find an accurate representation of  $g$. A popular approach for low-dimensional systems when derivative data $dh/dt$ is available is the `Sparse Identification of Nonlinear Dynamics (SINDy)' \cite{sindy,pde_sindy}. However, SINDy \CRCAC{requires a library of terms and struggles with } 
complex chaotic dynamical systems as shown by \cite{liu2024data}.
In this work, we will apply a more general framework, known as `neural ODEs' (NODE) \cite{Chen2019}, in which 
\AL{we approximate the vector field in the manifold coordinates ($g$) directly with a neural network.}
Its efficacy in forecasting chaotic fluid dynamics has been demonstrated in \cite{alec_chaos,alec_coutte,Chakraborty}. \AL{NODEs have the advantages of being Markovian and continuous in time.}

Next, we focus on describing the state-of-the-art \AL{data-driven} prediction frameworks for chaotic dynamics in the full state space. When \CRCA{equations of motion are unknown, and} only temporal snapshots \CRCA{of data} are available, 
forecasting can be done using reservoir networks \cite{Pathak,Doan,Fan,Platt,Racca}, recurrent neural networks \cite{Vlachas}, \CRCA{transformers \cite{Geneva,Gilpin} and quantum reservoir computing \cite{ahmed2024}}. These methods use discrete time maps, they are non-Markovian, and \AL{often} increase the dimension of the state space \AL{gain a} \CRCA{richer feature representation of the true state of the system.} 
The latter becomes a problem when dealing with large spatial domains with chaotic dynamics as results in excessive computational demands. To tackle this challenge, \cite{Pathak} proposed a parallel approach: assigning distinct reservoir networks to various spatial domains. For the modeling of the 1D KSE \CRCA{with a domain size of $L=200$, the full state dimension of the discretized system was 500. Using } local data, the system was divided into 64 reservoirs, each with a reservoir dimension of 5000. Thus, the dimension of the representation has not been reduced, but rather expanded, by two orders of magnitude. Their framework faithfully captured the chaotic dynamics for multiple Lyapunov times.
\AL{Here, we emphasize, that our approach differs from \cite{Pathak} in that we dramatically \emph{reduce} the dimension of the state, and we find a Markovian representation of the dynamics.}
\CRCAC{Finally, \cite{Arcomano2022} used the local modeling approach developed in \citep{Pathak} for geophysical data. }

Past data-driven reduced-order modeling approaches take single  
manifold coordinates to globally represent the entire spatial domain \cite{Milano,Page_2021, alec_coutte,carlos}.
A global parametrization of the manifold becomes challenging as the state-space sizes increase, thereby limiting the extent of possible dimension reduction -- see \cite{bregler1993surface,roweis2001global}.
For example, it has been suggested that the manifold dimension for the Kuramoto-Sivashinsky equation (KSE) scales linearly with the domain length $L$. In this work, we will discuss a case with a domain size of $L=220$, which leads to an estimated manifold dimension of $d_h\approx100$ \cite{Yang_2009}. However, a global model with this dimension does not lead to accurate reconstruction (as we show in Section 3.A). In response to this challenge,  we propose a solution by breaking the state space into localized spatial regions, referred to as `patches', treating each as a distinct dynamical system that is equivalent  and communicates with others.
This approach involves creating local data-driven models that effectively capture the system's dynamics while significantly reducing the dimension of the data. \CRCAC{Each patch is very low-dimensional in comparison to \cite{Pathak}.} Subsequently, we learn the dynamics with this low-dimensional representation using NODEs, leveraging its capability to represent the dynamics as a continuous-time Markovian system. 
\CRCAC{Here, we treat each patch equivalently (e.g., we use the same coordinate transformation from the full state and the same vector field for the dynamics), which drastically improves training by reducing the number of snapshots required.}
We call this method `Distributed Data-driven Manifold Dynamics' (DisDManD), and use it to build ROMs for the chaotic KSE and 2D  Kolmogorov flow. 

\section{Framework} \label{sec:Framework}

\begin{figure*} 
    \centering
	\includegraphics[width=\textwidth,clip]{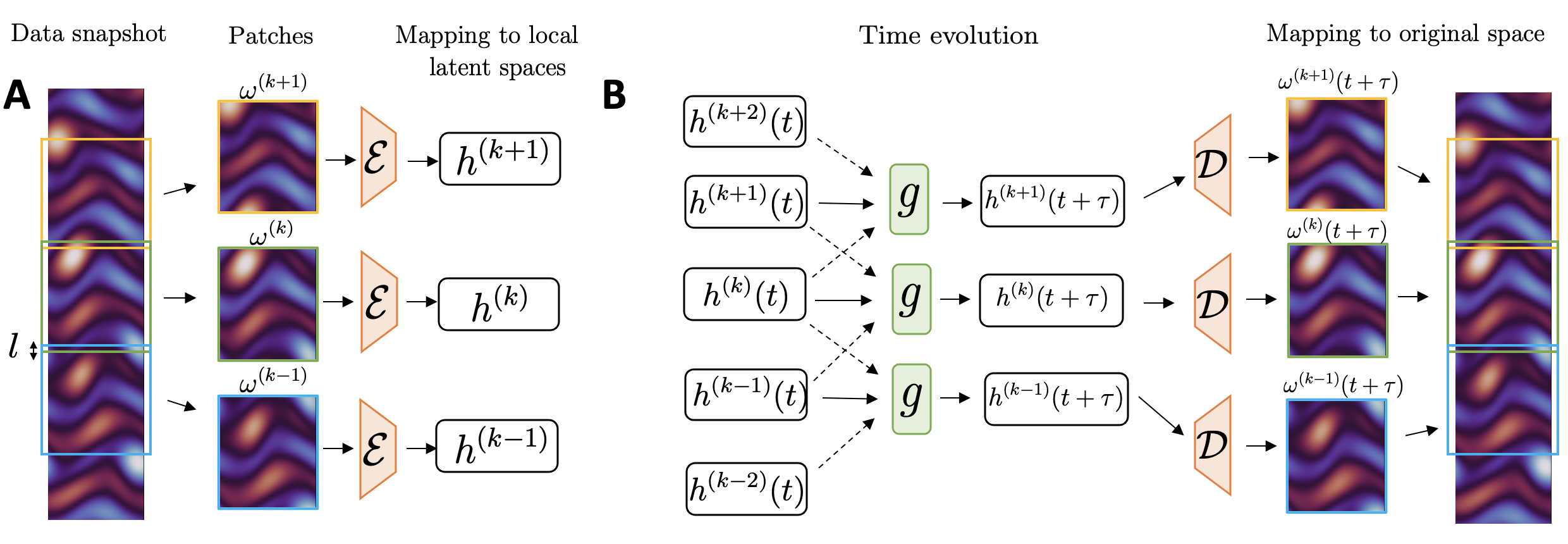}
	\caption{Overview of our approach applied to the vorticity fields for 2D Kolmogorov flow. (A)  The spatial domain is divided into patches denoted as $(k)$. A shared autoencoder maps these fields to a low-dimensional representation 
 $h^{(k)}=\mathcal{E}(\omega^{(k)})$, and its inverse to recover $\tilde{\omega}^{(k)}=\mathcal{D}(h^{(k)})$.
 (B) NODEs evolve the low-dimensional patches with vector field $g$ by incorporating interactions with neighboring patches (e.g., $h^{(k)}$, $h^{(\textit{neighbors})}=(h^{(k-1)},h^{(k+1)})$). 
 Upon completion of the time evolution, decoding is applied to reconstruct the full vorticity field from each patch.
    \label{framework}}
\end{figure*}


\AL{First, we generate training data for our method by forecasting the full state forward in time and sampling $N$ snapshots of data (e.g., the full state comes from a fully resolved direct numerical simulation).}
Then, we divide the domain into  $K$ `patches' of data. Instead of discovering a global representation of an invariant manifold for $u$ \CRCA{(e.g., $h_g=\mathcal{E}_g(u;\theta$))},
we search for many local low-dimensional representations for each patch
\begin{equation} \label{eq:patch1}
	h^{(k)}=\mathcal{E}(u^{(k)};\theta_\mathcal{E}),
\end{equation}
where $h^{(k)}\in\mathbb{R}^{d_k}$, along with its inverse 
\begin{equation} \label{eq:patch2}
	\tilde{u}^{(k)}=\mathcal{D}(h^{(k)};\theta_\mathcal{D}).
\end{equation}
\AL{The weights of the encoder $\mathcal{E}$ and decoder $\mathcal{D}$ are given by $\theta_\mathcal{E}$ and $\theta_\mathcal{D}$.} 
These functions can in principle reconstruct the state (i.e., $u^{(k)}=\tilde{u}^{(k)}$). \CRCAC{In this formulation, each patch has the same coordinate transformation to the low-dimensional representation. }
Once the low-dimensional representation of the \CRCA{state-space}  
in each patch has been discovered, \CRCA{we learn an evolution equation for each patch by using a (stabilised) NODE approach, described by}
\begin{equation} \label{eq:patch3a}
\dfrac{dh^{(k)}}{dt}=g(h^{(0)},...,h^{(K-1)};\theta_g) + Ah^{(k)}, ~~k=1,\cdot\cdot\cdot,K.
\end{equation}
\CRCA{the second term of the RHS corresponds to a stabilisation term that prevents solutions to drift away from the attractor \cite{alec_chaos,Floryan2024}. Here we set $A$ to be the  diagonal matrix $A=-\beta\delta_{ij}\sigma_ih^{(k)}$ where $\sigma_ih^{(k)}$ stands for the standard deviation of the ith component of $h$, $\beta$ is a fixed parameter and $\delta_{ij}$ is the Kronecker delta. }


If we assume only neighboring patches affect each other, we can rewrite Eq. \ref{eq:patch3a} as 
\begin{equation} \label{eq:patch3}
\dfrac{dh^{(k)}}{dt}=g(h^{(k)},h^{(\textit{neighbors})};\theta_g) + Ah^{(k)}.
\end{equation} 
To mitigate sharp changes at the domain boundaries, we overlap the patches by some length $l$, effectively expanding the region of each patch \AL{(here $l$ is the number of grid points)}. 
\AL{We then perform a weighted averaging of neighboring patches at the boundary resulting in continuity of the full state.}
\AL{Note, that this is effectively a weighted linear regression, which could lead to discontinuities in the derivative of the field. Here, we aimed for simplicity, but if smoother transitions between patches are required one could just as easily use polynomials or other nonlinear functions.}

We integrate Eq. \ref{eq:patch3} forward in time for all $K$ patches at once resulting in the prediction
\begin{equation}\label{eq:patchODE}
	\tilde{h}^{(k)}(t_i+\tau)=h^{(k)}(t_i)+\int_{t_i}^{t_i+\tau}g(h^{(k)}, h^{(\textit{neighbors})};\theta_g) + Ah^{(k)} dt,
\end{equation}
where $\tau$ represents the training forecast horizon.

\CRCAC{To find $\mathcal{E}$, $\mathcal{D}$}, we opted for standard autoencoders to generalize the framework, \CRCAC{while other more advanced autoencoders could be used \cite{alec_pre,irmae}. }
An autoencoder outputs $\tilde{u}^{(k)}=\mathcal{D}(\mathcal{E}(u^{(k)};\theta_\mathcal{E});\theta_\mathcal{D})$, the parameters of which we train to minimize 
\begin{equation} \label{eq:LossDistAuto}
	L=\dfrac{1}{dNK}\sum_{i=1}^N\sum_{k=1}^K||u^{(k)}(t_i)-\tilde{u}^{(k)}(t_i)||_2^2,
\end{equation}
\AL{where 
$\tilde{u}^{(k)}(t_i)$ is the reconstruction and $u^{(k)}(t_i)$ is the reference data.}
To learn the dynamics, we train a NODE 
to minimize the difference between the prediction 
$\tilde{h}^{(k)}(t_i+\tau)$ and the known data $h^{(k)}(t_i+\tau)$. This loss function is defined by
\begin{equation} \label{eq:LossDistNODE}
	J=\dfrac{1}{dNK}\sum_{i=1}^N\sum_{k=1}^K||h^{(k)}(t_i+\tau)-\tilde{h}^{(k)}(t_i+\tau)||_2^2.
\end{equation}
To determine the derivatives of $J$ with respect to the neural network parameters $\partial J/\partial\theta$, we use automatic differentiation  \cite{alec_coutte}. 
We optimize the previous loss functions using an Adam optimiser in PyTorch \citep{pythorch}.

Fig.\ \ref{framework} shows a schematic of our methodology applied to data obtained from a solution of 2D Kolmogorov flow, featuring only connections \CRCA{in the y-direction}.
In Fig. \ref{framework}A, we present the learning of the low-dimensional representation coordinates of each patch  (e.g., $h^{(k)}$ and its neighbor patches $h^{(k-1)}$ and $h^{(k+1)}$). 
Fig.\ \ref{framework}B illustrates the framework to learn 
 temporal evolution of coordinates $h^{(k)}$ given by vector field ${g}$; the temporal evolution of the coordinates $h^{(k)}$  also incorporates information from its neighbors. Once these functions are learned, new initial conditions can be mapped to the coordinates of the low-dimensional representation, evolved forward in time, and then mapped back to the full space.

\section{Results} \label{sec:Results}

We demonstrate our framework on three datasets with increasingly complex structures to show the predictive capabilities with respect to short-time tracking and long-time statistics. First, we consider the dynamics of the one-dimensional KSE, which is the simplest PDE that exhibits chaotic behavior \citep{Brummit2009}. Then, we extend our analysis to two-dimensional Kolmogorov flow, a system that captures characteristics of turbulent behavior in fluid dynamics. We consider Kolmogorov flow in a narrow domain where neighbors only exist in one direction, and we consider Kolmogorov flow in a large domain where neighbors exist in all directions.

\subsection{Kuramoto-Sivashinsky Equation} 
\label{sec:KSE}

The first case we consider is the one-dimensional KSE,
\begin{equation}
\frac{\partial  u}{\partial t}=  
-u\frac{\partial u }{\partial x} 
-\frac{\partial^2 u }{\partial x^2}
-\frac{\partial^4 u }{\partial x^4},
\end{equation}
for $x\in[-L/2, L/2]$, with periodic boundary conditions and \AL{domain size} $L=220$, which yields chaotic dynamics.
The negative sign on the second derivative of the KSE causes energy production, while the fourth derivative term causes dissipation \cite{chaos_book}.
To generate the dataset, we discretize the PDE using a pseudo-spectral method with $500$ Fourier modes, $u \in \mathbb{R}^{500}$, and assemble a dataset consisting of 
\CRCAC{8000 time units from a single trajectory with snapshots separated every $0.25$ time units.}
 \CRCAC{We split the data into 80\% training data and 20\% testing data.}
\CRCA{We dropped the first 1000 time units as transient data}
\AL{so that the data lies near the attractor}. For this case, we separate the dataset into $K=10$ patches with an overlap of $l=1$ grid points.
\AL{This choice was informed by the fact that low-order models accurately capture the chaotic dynamics of the KSE with a domain size of $L=22$ \citep{alec_pre}.}
Each patch has a dimension 
of $d_h=12$, resulting in a
\AL{ system with $K \cdot d_k=120$ degrees of freedom,}
as opposed to the original state space $u\in \mathbb{R}^{500}$.
\CRCAC{To highlight the advantages brought by splitting the full state space into patches, we also consider predictions with a global model with $h_g\in \mathbb{R}^{120}$, and referred to as `DManD'.
We chose a dimension of 120  to match  the dimension of DisManD (e.g., $Kd_h$).}

\begin{figure*}
\centering
\begin{tabular}{c}
\includegraphics[width=0.95\textwidth,clip]{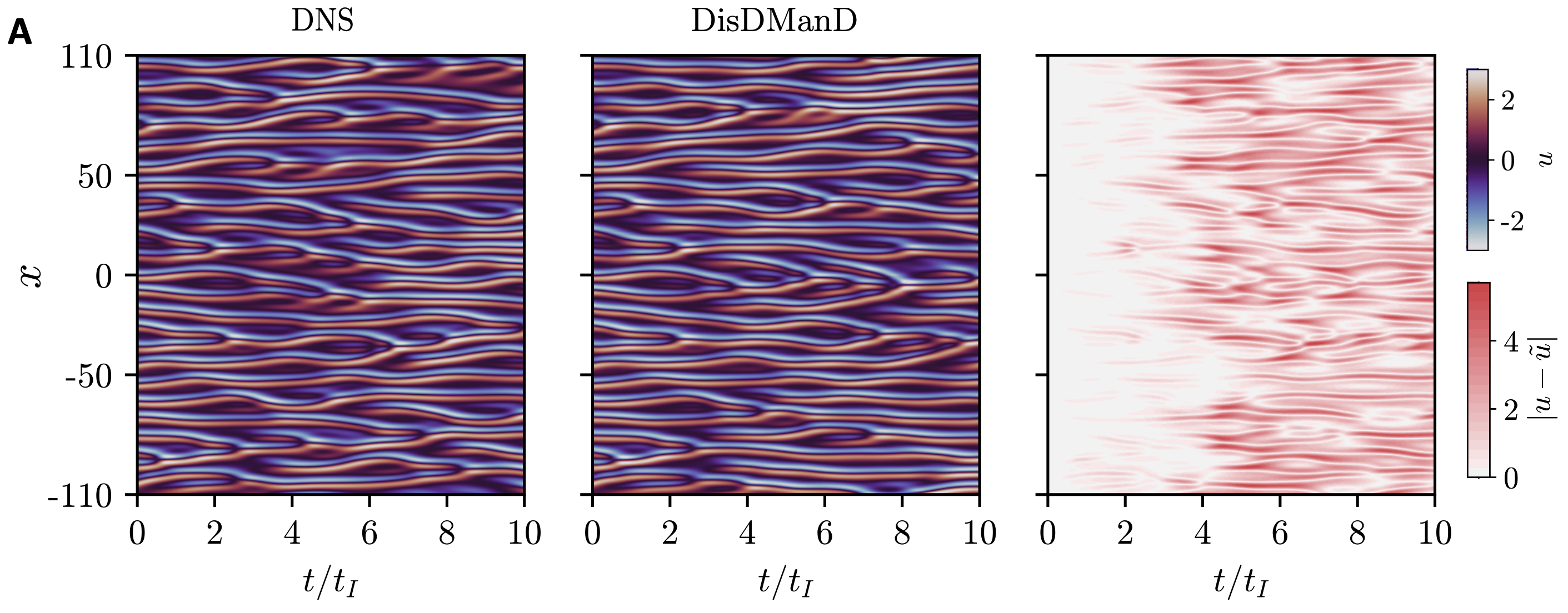} \\
\includegraphics[width=0.85\textwidth,clip]{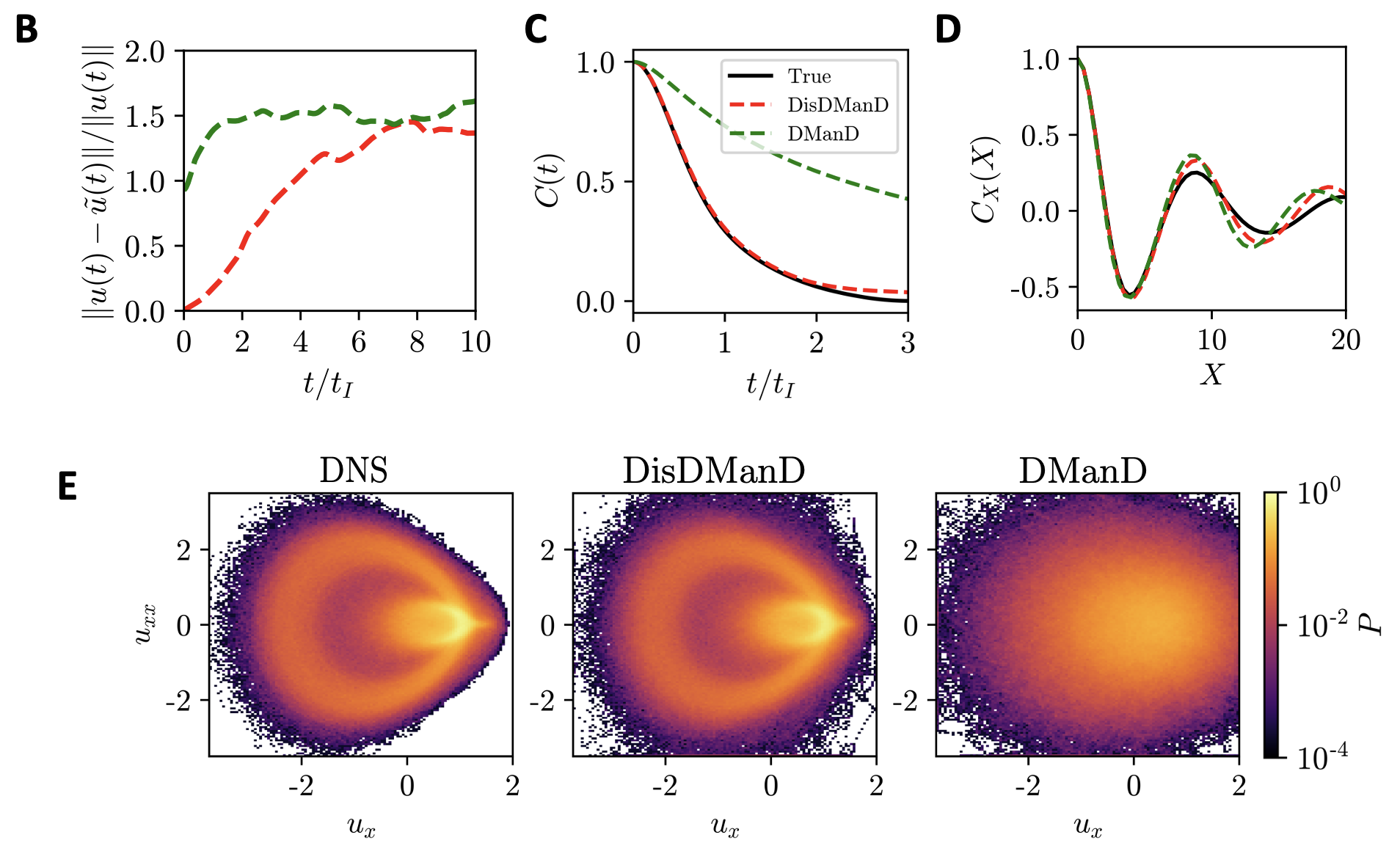} \\
\end{tabular}
\caption{ DisDManD and DManD for the KSE with domain size $L=220$.
(A) Left to right:   Representation of the \CRCA{dynamics}
for the true solution and DisDManD model using the same initial condition, and the error between the true and predicted trajectories up to $t/t_I=10$. 
(B) 
Ensemble-averaged short-time tracking for  DisDManD and DManD (with $Kd_h=120$).
(C)  Temporal autocorrelation function for the ground truth and the models. 
(D)  Spatial autocorrelation function. 
(E) Joint PDFs of $u_x$ and $u_{xx}$ for the true data, DisDManD and DManD, respectively.}
\label{KSE}
\end{figure*}

In Fig.\ \ref{KSE}A, we showcase the predictive capabilities of  DisDManD for the time evolution of the same initial condition with the DNS and the model (left and middle panels, respectively). The right panel illustrates the absolute error of the prediction $|u-\tilde{u}|$. Time is normalised by the characteristic integral time $t_I=\int_{0}^{\infty}C(t)dt\approx 7$, where $C$ corresponds to the temporal autocorrelation, defined as
\CRCA{$C(t)=\left \langle u(0) u(t) \right \rangle/\left \langle u(0)^2 \right \rangle$}.
Notably, there are no discernible discontinuities in the field (see right panel of Fig. \ref{KSE}A). This smooth continuity is a result of taking into consideration the neighboring effects while learning the evolution equation for each patch.

Initially, the model trajectories closely align with the ground truth for approximately $t/t_I  \approx 3$  before diverging, \CRCA{which indicates that  DisDManD framework effectively reproduces the true dynamics \CRCAC{for a short time}.}
Fig. \ref{KSE}B displays the ensemble relative tracking error $\left \|  u(t) - \tilde{u}(t) \right \|/ \left \|  u(t)  \right \|$ of 100 random trajectories at time $t_i$. 
We observe that a relative  error of unity is reached at $t/t_I\approx 5$, after which the curve \AL{plateaus, indicating the trajectories tend to diverge at this time.}
\CRCAC{The DManD model proves incapable of accurately predicting the short-term trajectories. Fig. \ref{KSE}B  highlights the poor performance of the autoencoder in finding the correct coordinate transformation from the full state to the low-dimensional representation as the relative error is already high at $t/t_I=0$. Therefore, learning the dynamics from a poor representation of the data will lead to inaccurate predictions. }
To further understand this behavior, Fig.\ \ref{KSE}C presents the temporal autocorrelation. 
\AL{The DisDManD model accurately captures the temporal autocorrelation, while the global model shows much higher correlation.}

\AL{Next, we evaluate the long-time statistical predictive capabilities of DisDManD.}
Fig. \ref{KSE}D shows the spatial autocorrelation function $C_X(X)=\langle{u(x,t)u(x+X,t)\rangle}/\langle{u(x,t)\rangle}$.
\AL{Both models accurately capture the spatial autocorrelation, indicating that the autoencoder in the DManD model outputs fields with the correct spatial autocorrelation, despite the fact that temporal modeling of the DManD model is poor.}
Fig.\ \ref{KSE}E compares the joint probability density function (PDFs) of the pointwise values of the first ($u_x$) and second  ($u_{xx}$) derivatives of the state evolved forward to $t/t_I=1000$ \CRCAC{from an initial condition that was started on the attractor}. It shows that DisDManD faithfully reconstructs the long-time dynamics of the system as it accurately captures the core and boundaries of the PDFs. \CRCAC{DManD also falls short in capturing the temporal autocorrelation  and exhibits poor performance in predicting the first ($u_x$) and second  ($u_{xx}$) derivatives over extended time periods (see right panel of Fig.  \ref{KSE}E). }

\CRCAC{Therefore, a global model with the same dimension as DisDManD is inadequate for faithfully representing the intricate dynamics of the system because the autoencoder does not perform a good mapping to manifold coordinates, even though the manifold dimension for $L=220$ is estimated to be around $d_\mathcal{M}\approx 100$. This underscores the limitations of using a global reduced-order model for temporal predictions in large spatial chaotic systems.}


\subsection{Kolmogorov Flow} \label{sec:KFlow}

Here, we show the generalization of this framework to a fluid flow satisfying the Navier-Stokes Equations (NSE) in a two-dimensional physical domains. 
We consider  2D turbulence with a 
(sinusoidal) body force, known as  Kolmogorov flow \cite{arnol1960kolmogorov}. 
First, we present a case in which \AL{we break the flow into patches in one direction (the y-direction). Then, we present a case in which we break the flow into patches in two directions (the x- and y-directions).}

\begin{figure*}[ht!]
\centering
\begin{tabular}{c}
\includegraphics[trim=0 0 0 0,width=\textwidth,clip]{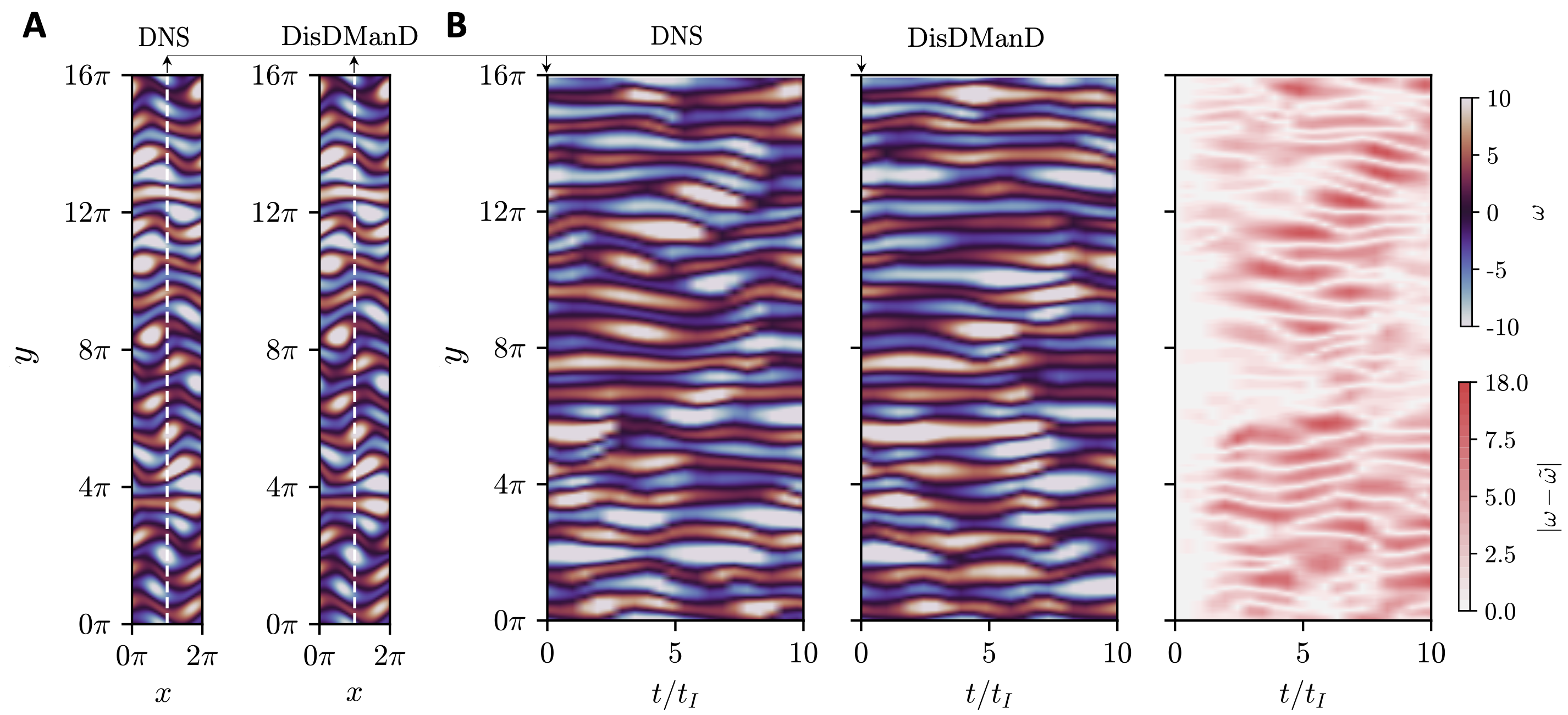} \\
\includegraphics[trim=0 0 0 0,width=0.8\textwidth,clip]{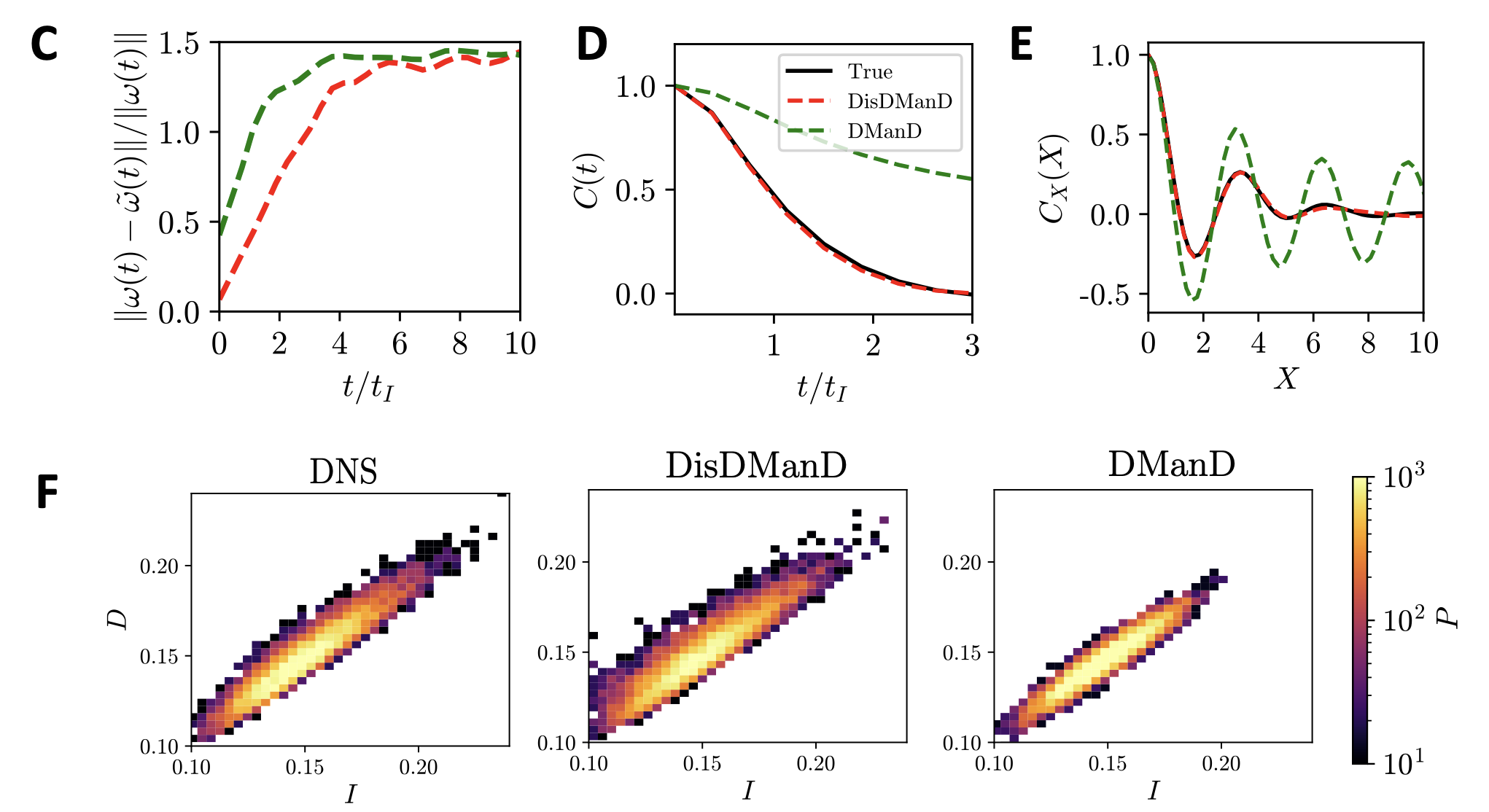} 
\end{tabular}
\caption{ 
DisDManD and DManD for 2D   Kolmogorov flow within a spatial domain of size  $[x,y] = [2\pi, 16\pi ]$ with $Re=200$.
(A) \CRCA{Reconstruction of} vorticity fields at $t/t_I=0$ 
for both the true data and DisDManD. 
(B) Temporal evolution of vorticity in a $(y,t)$ plane at a fixed $x=\pi$ (which corresponds to the dashed lines of panel A) up to $t/t_I=10$ for the true solution, DisDManD, and the absolute error between them, corresponding to the left, middle, and right panels, respectively. (C) 
Ensemble-averaged short-time tracking for   DisDManD and DManD (with $d_h=120$), respectively.  (D)  Temporal autocorrelation for DManD and DisDManD.
(E)  Spatial autocorrelation function in an $(y,t)$ plane at a fixed $x=\pi$ (which corresponds to the dashed lines of panel A). 
(F)  
Joint PDFs of Dissipation-Input for the true data, DisDManD and DManD, respectively.
\label{figure_tileY}
}
\end{figure*}

\begin{figure*}[ht!]
\centering
\begin{tabular}{cc}
\includegraphics[trim=0 0 0 0,width=\textwidth,clip]{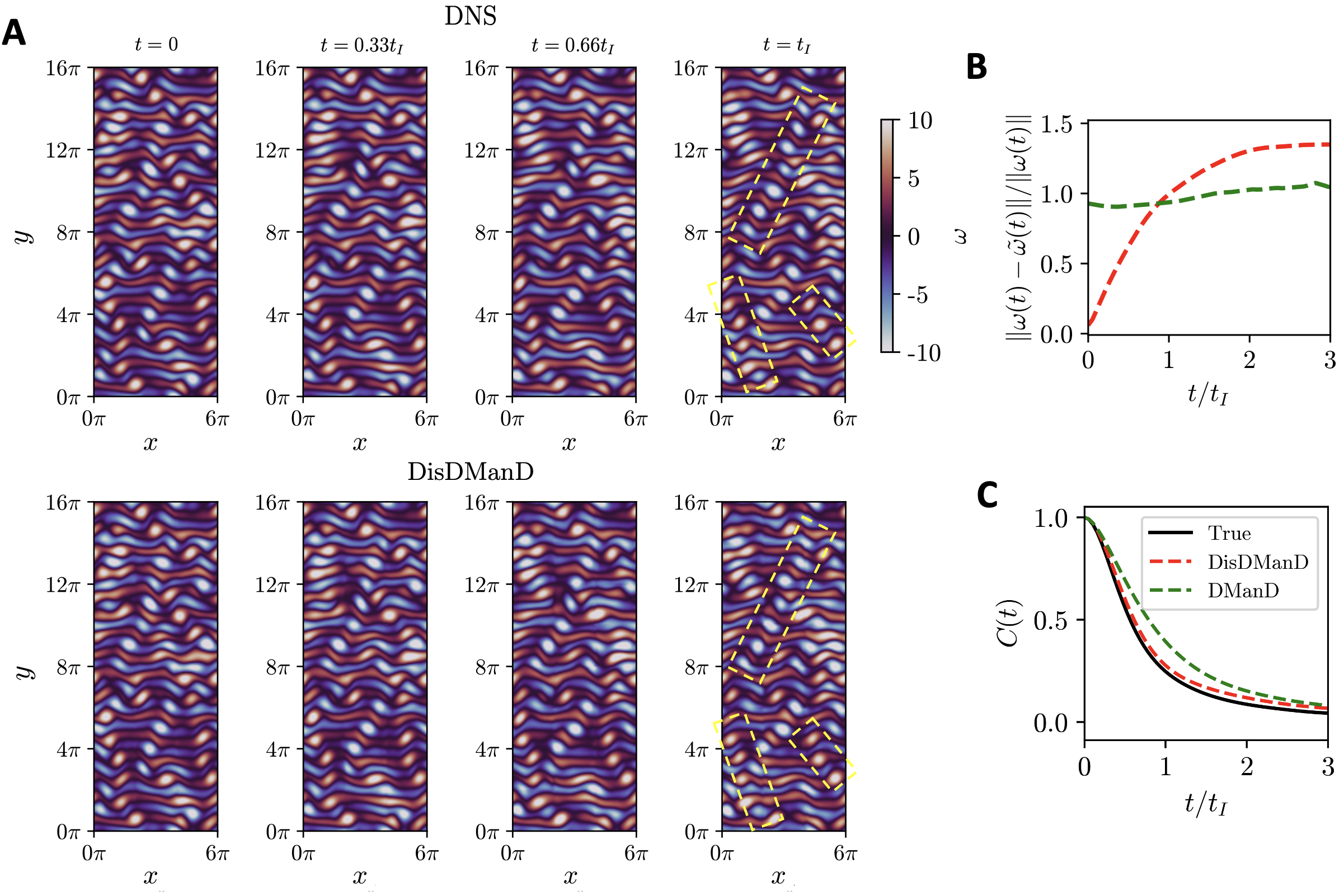} &
\end{tabular}
\begin{tabular}{c}
\includegraphics[trim=0 0 0 0,width=\textwidth,clip]{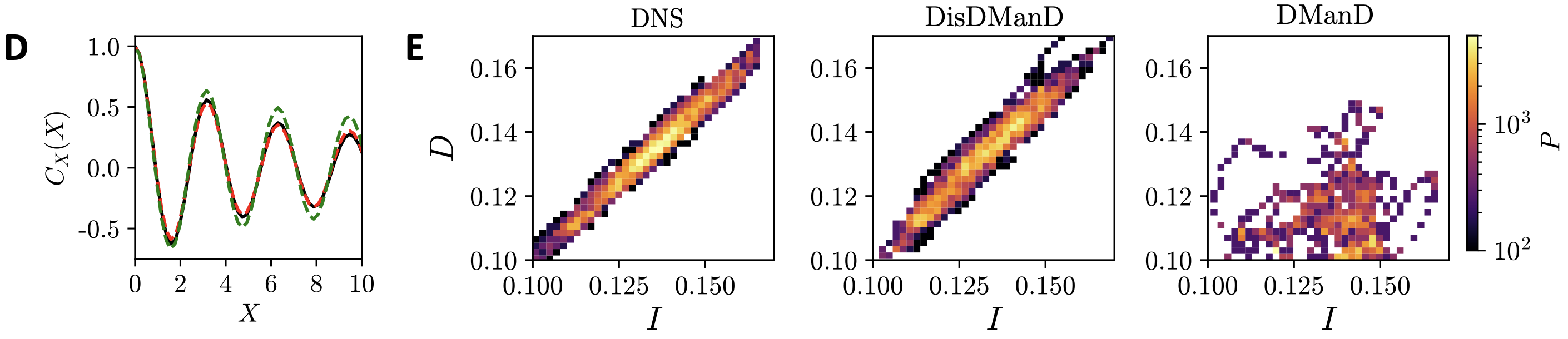} \\
\end{tabular}
 \caption{
DisDManD and DManD for 2D   Kolmogorov flow within a spatial domain of size  $[x,y] = [6\pi, 16\pi ]$ with $Re=200$.
 (A) \CRCA{Vorticity fields for DNS and DisDManD correspond to top and bottom panels, respectively. Normalised time is shown in each panel of the DNS. 
 The yellow dashed boxes indicate the  alignment of vortical structures  in the state space.}
(B) Ensemble-averaged short-time tracking for DisDManD and  DManD (with $Kd_h=720$), respectively.  (C)  Temporal autocorrelation for true and the models. 
(D) Spatial autorrelation in an $(y,t)$ plane at a fixed $x=3\pi$.
(E) Joint PDFs of Dissipation-Input for the true data, DisDManD and DManD, respectively.
}\label{figure_tileXY}
\end{figure*}

Simulations were performed by solving the incompressible Navier–Stokes equations:
\begin{equation}\label{div}
 \nabla \cdot \textbf{u}=0, ~~~
 \frac{\partial \textbf{u}}{\partial t}+\textbf{u} \cdot\nabla \textbf{u} + \nabla p  =  + \frac{1}{Re}~   \nabla ^2 \textbf{u}  + \sin (n_f y) \textbf{e}_x 
\end{equation}
where  $t$ and  $\textbf{u}$, and $p$ stand for time, velocity, and pressure, respectively. The equations are solved in a doubly periodic domain of size $[0,L_x] \times [0, L_y]= [0, 2\pi] \times [0, 2\pi \alpha]$ (i.e., $\alpha=L_y/L_x$). The Reynolds number represents the balance between the convection and diffusion terms of equation \ref{div}, and is  given by  $Re=(\sqrt{\chi_f}/\nu)  (L_y/2\pi)^{3/2}$ where $n_f$, $\nu$, and $\chi_f$ represent the forcing wavenumber, the kinematic viscosity and the forcing amplitude per unit mass of the fluid, respectively.

Data was generated using the vorticity representation $\omega=\nabla \times \mathbf{u}$ with $\delta t =0.005$ following the pseudospectral scheme described by \cite{chandler_kerswell_2013}. For time-stepping, we made use of an implicit Crank-Nicholson scheme for the viscous terms. 
The numerical resolutions used  32 Fourier modes per $2\pi$. We choose each patch to cover a spatial domain of size $[L_x,L_y]=[2\pi, 2\pi]$, thus $u^{(k)}\in \mathbb{R}^{1024}$. This choice is motivated by the aim to cover the forcing of $n_f=2$ applied within that specific patch \CRCAC{(e.g., this patch size can be accurately captured, as shown by \cite{carlos})}.
Simulations were initialized from random divergence-free initial conditions, we dropped the early transient dynamics and selected $40\times 10^5$ snapshots of the flow field, separated by $\tau=0.5$ time units and $\tau=0.1$ for cases B.1 and B.2, respectively. \AL{We split the data into 80\% training data and 20\% testing data.}
The neural network training uses only the training data, and all comparisons use test data.

\subsubsection{ 
Domain size: $[L_x,L_y] = [2\pi, 16\pi ]$} 
First, we consider a case with $Re=200$, and an aspect ratio $\alpha=1/8$; thus,  the dimension of the full state is $\omega \in \mathbb{R}^{8,192}$. For this case, we split the state space in the y-direction into $K=8$ patches with an overlap of $l=1$ grid points.
To build the low-dimensional model, we assume that each patch has a dimension of  $d_h=15$ (as described in Supplementary Information),  resulting in a global dimension of \AL{$K\cdot d_k=120$ (nearly two orders of magnitude lower than the simulation).} 
\CRCAC{Finally,  we also compare our results with a global DManD model with $h\in \mathbb{R}^{120}$.}

The left and middle panels of Fig. \ref{figure_tileY}A display vorticity snapshots at $t/t_I=0$ for the true solution and the corresponding reconstruction with the autoencoders. We observe qualitatively that the low-dimensional representation can capture well the complex features of the system. Fig. \ref{figure_tileY}B depicts the temporal evolution of the vorticity field in a $(y,t)$-plane for $L_x=\pi$ up to $t/t_I=10$ for the   DNS and DisDManD for the same initial condition presented in panel A of figure \ref{figure_tileY} (here $t_I= 1.32$). The right panel in Fig. \ref{figure_tileY}B presents the error, $|  \omega(\pi,t) - \tilde{\omega}(\pi,t)|$. \AL{ The true dynamics are in excellent quantitative agreement for approximately an integral time $t/t_I \approx 1$, after which the trajectories diverge. Due to the turbulent nature of the system, we only expect the model to capture on the order of one integral time, especially considering the significant dimension reduction we perform. Furthermore, the trajectories show good qualitative agreement, which should lead to accurate long-time statistics.}

Fig.\ \ref{figure_tileY}C displays the  ensemble-averaged relative tracking error $\left \|  \omega(t) - \tilde{\omega}(t) \right \|/\left \|  \omega(t) \right \|$ of 
100 random trajectories. The tracking error rapidly rises until time $t/t_I \approx 4$ in which the error levels off. 
\CRCAC{For the DManD model, the autoencoders with the same global dimension as the DisDMaD  can not find an accurate mapping  from the full space  to the low-dimensional  representation as the relative reconstruction error at $t/t_I=0$ is around 0.5. This bad reconstruction of the data will subsequently lead to bad performance in the dynamics.}
In Fig. \ref{figure_tileY}D, the temporal autocorrelation $C(t)$ shows that the temporal horizon for the eventual loss of the correlation with respect to the initial conditions is around $t/t_I \approx 3$, and our DisDManD closely follows the temporal autocorrelation of the true dynamics (unlike the global DManD model). 
Fig. \ref{figure_tileY}E shows the spatial autocorrelation function $C_X(X)=\langle{\omega(x,t)\omega(x+X,t)\rangle}/\langle{\omega(x,t)\rangle}$, \CRCAC{we note that DisDManD accurately output vorticity fields with the correct spatial autocorrelation, while the DManD model behaves poorly.}
%

Next, we turn attention to the capabilities of DisDManD to reproduce key long-time properties of the system \AL{-- in particular we consider the energy balance.}
\CRCAC{ For Kolmogorov flow, the
dissipation rate and power input of this system are given by $\it{D}= \left \langle | \nabla \textbf{u}^{2}|\right \rangle_V /Re$ and  $\it{I}=  \left \langle  u \sin (n_fy) \right \rangle_V$, where $\langle \rangle_V  = \alpha/(4\pi^2)\int_0^{2\pi}\int_0^{2\pi/\alpha}dxdy$ is the volume average. 
}
Fig.\ \ref{figure_tileY}F displays the joint \AL{PDF of dissipation and power input}
for the DNS and  DisDManD from a single long trajectory up to  $t/t_I\approx3000$. 
DisDManD successfully captures the PDF of the true dynamics, exhibiting good agreement in the core of the input-dissipation relationship. Notably, the model aligns along the diagonal $D=I$, corresponding to dissipation balancing the energy input. 
\CRCA{The global DManD model overpredicts the input-dissipation core and underpredicts the rare events at the boundaries.}

\subsubsection{Domain size: $[L_x,L_y] = [6\pi, 16\pi ]$}

\AL{For our last case, we show the effectiveness of the framework for a larger spatial domain.}
We keep the same Reynolds number at $Re=200$ and extend the x-direction to an aspect ratio $\alpha=3/8$. This increases the state dimension of the full state $\omega \in \mathbb{R}^{24,576}$. \AL{At this larger aspect ratio, the dynamics in the x-direction become more complex, which requires us to now segment the domain in both the x- and y-direction.}
We split the domain into $K=24$ patches of size  $[2\pi, 2\pi]$ with an overlap of $l=1$ grid points. 
To build the low-dimensional model, we assume a local dimension of $d_h=30$ (as described in Supplementary Information), resulting in a global dimension of $K\cdot d_k=720$
-- a reduction by two orders of magnitude. 
As in the previous examples, we also build a DManD model with $h\in \mathbb{R}^{720}$.

Fig.\ \ref{figure_tileXY}A shows snapshots of the vorticity field for the DNS and DisManD up to one \AL{integral time ($t_I=1.7$)} \CRCA{ from an initial condition that was started on the attractor}. First, we observe a   good reconstruction from the autoencoders  at $t=0$. These snapshots reveal intricate multiscale features,  \CRCAC{such as  the presence of multiple vortex cores  aligned  with a length equal to half of the domain.
As time evolves, we observe a strong alignment of strong regions of vorticity under one integral time. At $t=t_I$, the DisDManD model is capable of predicting the location of the vortices in similar regions of the space as predicted by the DNS.}

To further illustrate the performance of the models with respect to the  short time tracking, Fig.\ \ref{figure_tileXY}B displays  the relative ensemble-averaged tracking error $\left \|  \omega(t) - \tilde{\omega}(t) \right \|/\left \|  \omega(t)  \right \|$ 
of 100 trajectories. The tracking error experiences a rapid increase until approximately $t/t_I\approx 2$, after which it levels off.
This early divergence from the initial condition is attributed to the highly chaotic nature of the system.
To support this observation, Fig.\ \ref{figure_tileXY}C shows the temporal autocorrelation $C(t)=\left \langle \omega(0) \omega(t) \right \rangle/\left \langle \omega(0)^2 \right \rangle$ of the vorticity field.
The temporal horizon for the eventual loss of correlation with respect to the initial conditions is around $t/t_I \approx 2$. Beyond this timescale, DisDManD is not expected to predict accurately due to the chaotic nature of the system.
\CRCA{ Fig.\ \ref{figure_tileXY}D shows the spatial autocorrelation $C_X(X)$ for the vorticity field along the centerline. This figure shows an excellent agreement between the DisDManD and the DNS, indicating that the model can accurately capture the features of the spatially large system. Finally, we note that DManD struggles to identify an effective coordinate transformation from the full state space to a low-dimensional representation, as evidenced by the normalized reconstruction error of approximately 0.8 at $t=0$ (see Fig. \ref{figure_tileXY}B).
Due to this inaccurate low-dimensional representation of the dynamics, the learned model fails to predict the true behavior of the system, as reflected in the poor performance of the ensemble-averaged tracking error.
\CRCA{For DManD, the autoencoders reconstruct the flow field as alternating vorticity bands in the x-direction. Specifically, the reconstruction fails to identify vortical cores advecting in the x-direction. This results in misleadingly accurate spatial and temporal autocorrelations, as the model does not accurately capture the underlying flow dynamics.)}
Additionally, the computational cost} \AL{of training the global models is substantially higher than DisDManD because far more weights are needed in the global model autoencoder.}

In Fig.\ \ref{figure_tileXY}E, we assess the long-term predictive capabilities of DisDManD and DManD model by plotting the joint PDF of dissipation and power input from a single extended trajectory spanning $t/t_I \approx 3\times10^2$. DisDManD  faithfully captures the PDF of the true dynamics, with a notable agreement in the core of the input-dissipation distribution. \AL{The prediction of the DManD model quickly diverges from the attractor, leading to an extremely poor reconstruction of the joint PDF.}

\section{Conclusions} \label{sec:Conclusion}

In this work, we presented a data-driven framework for building low-dimensional models for chaotic systems in large spatial domains.
\AL{This framework consists of segmenting a large domain into a set of patches, and then building models for each of the patches.}
\AL{These models consist of autoencoders, to dramatically reduce the dimension (here by two orders of magnitude), and neural ODEs, to evolve the dynamics with this low-dimensional representation. To accurately model the dynamics, we accounted for the influence of nearby patches in the model, and we developed an averaging method to enforce a continuous transition between patches at their boundaries.}
We applied this methodlogy to 1D KSE with $L=200$ and 2D Kolmogorov flow with $Re=200$ (at multiple aspect ratios). The resulting low-dimensional data-driven models effectively capture short-time dynamics over the span of one integral time, as well as long-time statistics such as the energy balance in the 2D Kolmogorov flow. We also demonstrate that a global model fails to capture the dynamics of these flows, highlighting the advantages of the framework presented here.

\CRCAC{In summary, this research makes a valuable contribution to the modeling of chaotic systems in large spatial domains. By focusing on local domain modeling, we can capture interactions between different scales while significantly reducing the dimensionality of the problem. For future work, we aim to apply this DisDManD framework to turbulent wall-bounded flows in larger domains, such as plane Couette flow or pipe flow, where the DManD framework has already been applied to study minimum domain sizes that sustain turbulence \cite{alec_coutte,pipe_jfm,kdmand}. Accurately modeling these larger systems could also facilitate the discovery of new Exact Coherent Structures (ECS), which have well-defined structures that organize the dynamics of chaotic flows, making their identification important, but particularly challenging in larger domains due to the high dimensional of the full state data.
}

\subsection*{Data Availability}
Source  code  for  our  models,  including  learned  components,  and  training  and  evaluation  datasets  are  available on GitHub

\subsection*{Acknowledgments}

This work was supported by ONR N00014-18-1-2865 (Vannevar Bush Faculty Fellowship).\\


\bibliography{main}

\end{document}

%% file: MDGcommands.tex
\newlength{\figwidth}
\newlength{\SCwidth}
\def\XXint#1#2#3{{\setbox0=\hbox{$#1{#2#3}{\int}$}
     \vcenter{\hbox{$#2#3$}}\kern-.5\wd0}}